\documentstyle[aps,pra,psfig,twocolumn]{revtex}
\begin{document}
\title{Thermal and quantum noise in active systems}
\author{Jean-Michel Courty\thanks{%
courty@spectro.jussieu.fr}, Francesca Grassia \thanks{%
grassia@spectro.jussieu.fr} and Serge Reynaud\thanks{%
reynaud@spectro.jussieu.fr}}
\address{Laboratoire Kastler Brossel\thanks{%
Unit\'{e} mixte de recherche de l'Universit\'{e} Pierre et Marie Curie, de
l'Ecole Normale Sup\'{e}rieure et du Centre National de la Recherche
Scientifique}\thanks{%
website: www.spectro.jussieu.fr/Mesure}, Case 74, 4 place Jussieu,\\
 F-75252 Paris Cedex 05, France}
\maketitle

\begin{abstract}
We present a quantum network approach to the treatment of thermal and
quantum fluctuations in measurement devices. The measurement is described as
a scattering process of input fluctuations towards output ones. We present
the results obtained with this method for the treatment of a cold damped
capacitive accelerometer.
\end{abstract}

\section{Non ideal quantum measurements}

Active systems are fundamental elements in high precision measurements.
Amplifiers are used either for amplifying the signal to a macroscopic level
or to make the system work around its optimal working point with the help of
feedback loops. With techniques such as cold damping, it is possible to
manipulate actively the fluctuations and to reduce the effective noise
temperature of the devices well below the operating temperature. The
analysis of sensitivity limits in these devices rises many questions related
to fundamental processes as well as experimental constraints. How far is it
possible to reduce the measurement temperature? How are these process
related to the fluctuation dissipation theorem? Are there quantum limits to
this noise reduction associated with Heisenberg inequalities? How do the
experimental constraints interplay with the fundamental limitations of the
sensitivity?

The aim of the present paper is to address these questions with quantum
network theory. This approach provides a rigorous thermodynamical framework
able to withstand the constraints of a quantum analysis of the measurement.
In the same time, it makes possible a realistic description of real
measurement devices. \ Thermodynamic and quantum fluctuations are treated in
the same footing. The measurement process is described as a scattering
process allowing for a modular analysis of real quantum systems. Active
systems such as the linear amplifier or the ideal operational amplifier are
described in this framework. Here, the approach will be illustrated by
analyzing the sensitivity of a cold damped capacitive accelerometer
developed for fundamental physics applications in space \cite
{Bernard91,Touboul92,Willemenot97}.

We first present the analysis of passive electrical systems in term of
quantum networks. Then, we use this approach to present the quantum analysis
of an operational amplifier working in the ideal limit of infinite gain,
infinite input impedance and null output impedance. In the last section, we
illustrate the theoretical framework with the example of a cold damped
accelerometer.

\section{Coupling with the environment}

Relations between fluctuations and dissipation have first been discovered by
Einstein who studied the viscous damping of mechanical systems \cite
{Einstein05}. Another important application was the study of Johnson-Nyquist
noise in resistive electrical elements \cite{Nyquist28}. This classical
result was extended to take into account the quantum statistical properties
of fluctuations\cite{Callen51,Landau84}. A general approach of these
relations was widely studied in the framework of linear response theory \cite
{Kubo66,Landau}.

\subsection{Dissipation and Fluctuations}

A first insight into the physical effect of the coupling of an electrical
circuit to the environment is provided by the analysis of an antenna in an
electrical resonator. When a current flows through the antenna,
electromagnetic radiation is emitted and the resonator energy decreases. As
far as the electric circuit is concerned, the effect of the antenna is the
same as a resistance. The antenna is also able to detect electromagnetic
fields. An incoming wave puts into motion the electrons in the antenna and
causes an electrical current to flow in the circuit. For thermal radiation,
the detection radiation leads to a random current which brings the
electrical oscillator to thermal equilibrium. In the high temperature limit,
it leads to the usual thermodynamic ${\frac12} k_{B}T$ per degree of
freedom, with $k_{B}$ being Boltzmann constant and $T$ the radiation
temperature. In the zero temperature limit, the detected field corresponds
to the vacuum fluctuations of the electromagnetic field and the induced
energy of the oscillator is the zero point energy ${\frac12} \hbar \omega
_{0}$, with $\omega _{0}$ the resonance frequency of the oscillator. 

\begin{figure}
\centerline{\psfig{figure=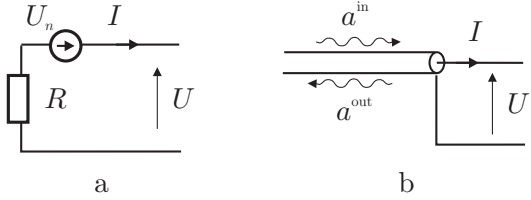,width=7cm}}
\vspace{2mm}
\caption{Representations of a resistance $R$. (a) Thevenin representation
with a voltage noise generator $U_{n}$. (b) Model with a semiinfinite line
and propagating fields $a^{in}$ and $a^{out}$}
\label{Figure1}
\end{figure}

In figure \ref{Figure1} are depicted two representations for a resistance $R$%
. Figure \ref{Figure1}a corresponds to the Thevenin representation with the
noise source represented as a voltage noise generator $U_{n}$. The relation
between the current $I$ and the voltage $V$ is 
\begin{equation}
U=RI+U_{n}  \label{resisthevenin}
\end{equation}
Figure \ref{Figure1} b corresponds to a model that originates from Nyquist's
analysis \cite{Nyquist28}. It consists in a semi infinite coaxial line of
characteristic impedance $R.$ The solution of the propagation equations in
the line may be written as the sum of two counterpropagating fields $I^{{\rm %
out}}$and $I^{{\rm in}}$ 
\begin{eqnarray}
I\left( x,t\right) &=&I^{{\rm out}}\left( t+\frac{x}{c}\right) -I^{{\rm in}%
}\left( t-\frac{x}{c}\right)  \nonumber \\
U\left( x,t\right) &=&R\left( I^{{\rm out}}\left( t+\frac{x}{c}\right) +I^{%
{\rm in}}\left( t-\frac{x}{c}\right) \right)
\end{eqnarray}
At the end of the line, we deduce the following relations: 
\begin{eqnarray}
&&U=RI+2RI^{{\rm in}}=RI+U_{n}  \nonumber \\
&&I^{{\rm out}}=I+I^{{\rm in}}
\end{eqnarray}
The first equation corresponds to the relation (\ref{resisthevenin}) and
leads to the identification of the noise as the input current $I^{{\rm in}}$%
. The second equation describes the output fields $I^{{\rm out}}$ emitted
back to the line.\ This output field may be used either to feed other
elements of the system or to perform a measurement by extracting information
on the system of interest through a line considered as the detection channel.

\subsection{Treatment with quantum fields}

In an infinite line, current and voltage may be treated as quantum fields
propagating in a two dimensional space-time. Throughout the paper, we will
consider that a function $f$ is defined in the time domain (notation $%
f\left( t\right) $) or in the frequency domain (Kubo's notation $f\left[
\omega \right] $) and that these two representations are related trough the
Fourier transform with the convention of quantum mechanics 
\begin{equation}
f\left( t\right) =\int \frac{d\omega }{2\pi }f\left[ \omega \right]
e^{-i\omega t}
\end{equation}
The electronics convention may be recovered by substituting $j$ to $-i$.

Free field operators $a^{{\rm in}}$ and $a^{{\rm out}}$ can be defined as
the Fourier components of $I^{{\rm in}}$and $I^{{\rm out}}$ 
\begin{eqnarray}
I\left( x,t\right)  &=&\int_{-\infty }^{\infty }\frac{d\omega }{2\pi }\sqrt{%
\frac{\hbar \left| \omega \right| }{2R}}\left( a^{{\rm out}}\left[ \omega %
\right] \exp \left[ -i\omega \left( t+\frac{x}{c}\right) \right] \right.  
\nonumber \\
&&-\left. a^{{\rm in}}\left[ \omega \right] \exp \left[ -i\omega \left( t-%
\frac{x}{c}\right) \right] \right)   \nonumber \\
U\left( x,t\right)  &=&\int_{-\infty }^{\infty }\frac{d\omega }{2\pi }\sqrt{%
\frac{\hbar \left| \omega \right| R}{2}}\left( a^{{\rm out}}\left[ \omega %
\right] \exp \left[ -i\omega \left( t+\frac{x}{c}\right) \right] \right.  
\nonumber \\
&&+\left. a^{{\rm in}}\left[ \omega \right] \exp \left[ -i\omega \left( t-%
\frac{x}{c}\right) \right] \right) 
\end{eqnarray}
They are normalized so that they obey the standard commutation relations 
\begin{eqnarray}
\left[ a^{{\rm in}}\left[ \omega \right] ,a^{{\rm in}}\left[ \omega ^{\prime
}\right] \right]  &=&\left[ a^{{\rm out}}\left[ \omega \right] ,a^{{\rm out}}%
\left[ \omega ^{\prime }\right] \right]   \nonumber \\
&=&2\pi \ \delta \left( \omega +\omega ^{\prime }\right) \ \varepsilon
\left( \omega \right)   \label{commutfree}
\end{eqnarray}
where $\varepsilon \left( \omega \right) $ denotes the sign of the frequency 
$\omega $. This relation just means that the positive and negative frequency
components correspond respectively to the annihilation $a_{\omega }$ and
creation $a_{\omega }^{\dagger }$ operators of quantum field theory 
\begin{equation}
a^{{\rm in}}\left[ \omega \right] =a_{\omega }\theta \left( \omega \right)
+a_{-\omega }^{\dagger }\theta \left( -\omega \right) 
\end{equation}
$\theta \left( \omega \right) $ denotes the Heavyside function.

To characterize the fluctuations of these noncommuting operators, we use the
correlation function defined as the average value of the symmetrized
product. With stationary noise, the correlation function depends only on the
time difference 
\begin{eqnarray}
\left\langle a^{{\rm in}}\left( t\right) \cdot a^{{\rm in}}\left( t^{\prime
}\right) \right\rangle &=&\sigma _{aa}^{{\rm in}}\left( t-t^{\prime }\right)
\nonumber \\
\left\langle a^{{\rm in}}\left[ \omega \right] \cdot a^{{\rm in}}\left[
\omega ^{\prime }\right] \right\rangle &=&2\pi \ \delta \left( \omega
+\omega ^{\prime }\right) \ \sigma _{aa}^{{\rm in}}\left[ \omega \right]
\end{eqnarray}
The dot symbol denotes a symmetrized product for quantum operators.

In the case of a thermal bath, the noise spectrum is 
\begin{equation}
\sigma _{aa}^{{\rm in}}\left[ \omega \right] =\frac{1}{\exp \frac{\hbar
\left| \omega \right| }{k_{B}T_{a}}-1}+\frac{1}{2}=\frac{1}{2}\coth \frac{%
\hbar \left| \omega \right| }{2k_{B}T_{a}}
\end{equation}
One recognizes the black body spectrum or the number of bosons per mode for
a field at temperature $T_{a}$ and a term $\frac{1}{2}$ corresponding to the
quantum fluctuations. The energy per mode will be denoted in the following
as an effective temperature $\Theta _{a}$ 
\begin{equation}
k_{B}\Theta _{a}=\hbar \left| \omega \right| \sigma _{aa}^{{\rm in}}\left[
\omega \right] =\frac{\hbar \left| \omega \right| }{2}\coth \frac{\hbar
\left| \omega \right| }{2k_{B}T_{a}}
\end{equation}
In the high temperature limit the classical energy for an harmonic field of $%
k_{B}T_{a}$ per mode is recovered. In the low temperature limit, the energy $%
\frac{\hbar \left| \omega \right| }{2}$ corresponding to the ground state of
a quantum harmonic oscillator is obtained. Note that the term $\frac{1}{2}$
corresponding to the zero point quantum fluctuations was added by Planck so
that the difference with the classical result $k_{B}T_{a}$ tends to zero in
the high temperature limit \cite{Planck}.

These results are easily translated to obtain the expression of the Johnson
Nyquist noise power 
\begin{equation}
\sigma _{U_{n}U_{n}}\left[ \omega \right] =2R\hbar \left| \omega \right|
\sigma _{aa}^{{\rm in}}\left[ \omega \right] =2Rk_{B}\Theta _{a}
\end{equation}
Our symmetric definition of the noise power spectrum leads to a factor $2$
difference with the electronic convention where only positive frequencies
are considered.

\subsection{Quantum networks}

The elementary systems described up to now as well as more complex devices
to be studied later in this paper may be described by using a systematic
approach which may be termed as ``quantum network theory''. Initially
designed as a quantum extension of the classical theory of electrical
networks \cite{Meixner63}, this theory was mainly developed through
applications to optical systems \cite{Yurke84,Gardiner88}. It has also been
viewed as a generalized quantum extension of the linear response theory
which is of interest for electrical systems as well \cite{Courty92}. It is
fruitful for analyzing non-ideal quantum measurements containing active
elements \cite{Francesca98,Grassia99}.

In this quantum network approach, the various fluctuations entering the
system, either by dissipative or by active elements, are described as input
fields in a number of lines as depicted on \ref{Figure1} b.

\begin{figure}[htb]
\centerline{\psfig{figure=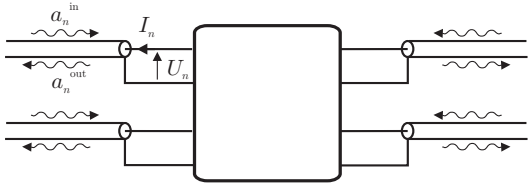,width=7cm}}
\vspace{4mm}

\caption{Representation of an electrical circuit as a quantum network. The
central box is a reactive multipole which connects noise lines corresponding
to the fluctuations entering the system, either by dissipative or by active
elements. For example, the upper left port $n$ with voltage $U_{n}$ and
current $I_{n}$ is connected to a line of impedance $R_{n}$ with inward and
outward fields $a_{n}^{{\rm in}}$ and $a_{n}^{{\rm out}}$.}
\label{Figure2}
\end{figure}

We first consider a passive linear network built with resistances and
reactive elements like capacitances or inductances. Each resistance $R_{n}$
is modeled as a semi-infinite coaxial line $a_{n}$ with characteristic
impedance $R_{n}$. The voltage $U_{n}$ and current $I_{n}$ associated with
the resistance are the inward and outward fields $a_{n}^{{\rm in}}$ and $%
a_{n}^{{\rm out}}$ evaluated at the end of this line 
\begin{eqnarray}
{\bf I} &=&{\bf R}^{-{\frac12} }\sqrt{\frac{\hbar \left| \omega \right| }{2}}%
\left( {\bf a}^{{\rm out}}-{\bf a}^{{\rm in}}\right)  \nonumber \\
{\bf U} &=&{\bf R}^{{\frac12} }\sqrt{\frac{\hbar \left| \omega \right| }{2}}%
\left( {\bf a}^{{\rm out}}+{\bf a}^{{\rm in}}\right)  \label{defnyquist}
\end{eqnarray}
Here, ${\bf X}={\bf I},{\bf U},{\bf a}^{{\rm in}},{\bf a}^{{\rm out}}$
denotes the column vector with components $X_{n}$ and ${\bf R}$ is the
diagonal matrix formed with the characteristic impedances $R_{n}$.

Input fields corresponding to different lines commute with each other. For
simplicity, we also consider that the fields entering through the various
ports are uncorrelated with each other. The interaction with the reactive
elements is described by a reactive impedance matrix ${\bf Z}$ 
\begin{eqnarray}
{\bf U} &=&-{\bf Z\ I}  \nonumber \\
{\bf Z^{\dagger }} &=&-{\bf Z}
\end{eqnarray}

The whole network is then associated with a scattering ${\bf S}$ matrix,
also called repartition matrix \cite{Feldmann}, describing the
transformation from the input fields to the output ones 
\begin{eqnarray}
&&{\bf a}^{{\rm out}}={\bf S\ a}^{{\rm in}}  \nonumber \\
&&{\bf S}=\frac{{\bf R}^{-{\frac12}}{\bf ZR}^{-{\frac12} }-{\bf 1}}{{\bf R}%
^{-{\frac12}}{\bf ZR}^{-{\frac12}}+{\bf 1}}
\end{eqnarray}
The output fields $a^{{\rm out}}$ are also free fields which obey the same
commutation relations \ref{commutfree} as the input ones. In other words, $%
{\bf S}$ matrix is unitary. In the case of the passive network, this
property is an immediate consequence of the reactive nature of the impedance
matrix ${\bf Z}$ 
\begin{equation}
{\bf S^{\dagger }}=\frac{{\bf R}^{-{\frac12}}{\bf Z^{\dagger }R}^{- {\frac12}%
}-{\bf 1}}{{\bf R}^{-{\frac12}}{\bf Z^{\dagger }R}^{- {\frac12}}+{\bf 1}}=%
\frac{-{\bf R}^{-{\frac12}}{\bf ZR}^{- {\frac12}}-{\bf 1}}{-{\bf R}^{-{\frac1%
2}}{\bf ZR}^{- {\frac12}}+{\bf 1}}={\bf S^{-1}}
\end{equation}

More generally, the unitarity of the ${\bf S}$ matrix is required to ensure
the quantum consistency of the description. In the following section, we
will make use of this property to deduce general properties of amplifiers.

\section{Fluctuations in amplifiers}

Quantum noise associated with linear amplifiers has been the subject of
numerous works. In the line of thought initiated by early works on
fluctuation-dissipation relations, active systems have been studied in the
optical domain when maser and laser amplifiers were developed \cite
{Heffner62,Haus62,Gordon63}. General thermodynamical constraints impose the
existence of fluctuations for amplification as well as dissipation
processes. The added noise determines the ultimate performance of linear
amplifiers \cite{Caves82,Loudon84} and plays a key role in the question of
optimal information transfer in optical communication systems \cite
{Gordon62,Takahasi65}.

We first consider the amplification of a field, for example in long distance
telecommunication systems with repeaters.

The amplification of the field $a^{{\rm in}}$ with a phase insensitive gain $%
G$ is given by the following equation 
\begin{equation}
a^{{\rm out}}=Ga^{{\rm in}}+B^{{\rm in}}
\end{equation}
where $B^{{\rm in}}$ is a noise added by the amplification. The gain $G$ may
be frequency dependent. The commutator of the output field is then 
\begin{eqnarray}
\left[ a^{{\rm out}}\left[ \omega \right] ,a^{{\rm out}}\left[ \omega
^{\prime }\right] \right]  &=&\left| G\right| ^{2}\left[ a^{{\rm in}}\left[
\omega \right] ,a^{{\rm in}}\left[ \omega ^{\prime }\right] \right]  
\nonumber \\
&&+\left[ B^{{\rm in}}\left[ \omega \right] ,B^{{\rm in}}\left[ \omega
^{\prime }\right] \right] 
\end{eqnarray}
The unitarity of the input output transformation and the preservation of the
commutation implies a non zero commutator for $B^{{\rm in}}$%
\begin{eqnarray}
\left[ B^{{\rm in}}\left[ \omega \right] ,B^{{\rm in}}\left[ \omega ^{\prime
}\right] \right]  &=&\left[ a^{{\rm in}}\left[ \omega \right] ,a^{{\rm in}}%
\left[ \omega ^{\prime }\right] \right] -\left[ a^{{\rm out}}\left[ \omega %
\right] ,a^{{\rm out}}\left[ \omega ^{\prime }\right] \right]   \nonumber \\
&=&\left( 1-\left| G\right| ^{2}\right) \ 2\pi \ \delta \left( \omega
+\omega ^{\prime }\right) \ \varepsilon \left( \omega \right) 
\end{eqnarray}
This result does not depend on the specific amplification process. For a
gain larger than unity, the added noise can be represented by a free field $%
b^{{\rm in}}$ with the usual commutation relation (\ref{commutfree}) 
\begin{equation}
B^{{\rm in}}\left[ \omega \right] =\sqrt{\left| G\right| ^{2}-1}\ b^{{\rm in}%
}\left[ -\omega \right] =\sqrt{\left| G\right| ^{2}-1}\left( b^{{\rm in}}%
\left[ \omega \right] \right) ^{\dagger }
\end{equation}
The presence of the conjugation is characteristic of amplification processes
and is encountered as soon as gain is present \cite{Caves82,Loudon84} .

We may use this example to describe the noise analysis in a measurement
process. Let us consider that the field $a^{{\rm in}}$ carries a signal $A $
superimposed with fluctuations $c^{{\rm in}}$%
\begin{equation}
A=\left\langle a^{{\rm in}}\right\rangle ,\qquad c^{{\rm in}}=a^{{\rm in}%
}-\left\langle a^{{\rm in}}\right\rangle
\end{equation}
The input noise power $\Sigma _{AA}^{{\rm in}}$ corresponds to the
fluctuations $\sigma _{cc}^{{\rm in}}$ 
\begin{equation}
\Sigma _{AA}^{{\rm in}}=\sigma _{cc}^{{\rm in}}
\end{equation}
The measurement corresponds to the output $a^{{\rm out}}$ of the amplifier 
\begin{equation}
a^{{\rm out}}=GA+Gc^{{\rm in}}+\sqrt{\left| G\right| ^{2}-1}\ b^{{\rm in}%
\dagger }
\end{equation}
To analyze the noise of this amplified signal, we define an estimator $%
\widehat{A}$ by normalizing the output field $a^{{\rm out}}$of the amplifier
so that it is the sum of $A$ and an extra noise 
\begin{equation}
\widehat{A}=\frac{1}{G}a^{{\rm out}}=A+c^{{\rm in}}+\sqrt{1-\frac{1}{\left|
G\right| ^{2}}}\ b^{{\rm in}\dagger }
\end{equation}
The added noise $\Sigma _{AA}^{{\rm out}}$ is then described by a spectrum 
\begin{equation}
\Sigma _{FF}^{{\rm out}}=\sigma _{cc}^{{\rm in}}+\left( 1-\frac{1}{\left|
G\right| ^{2}}\right) \sigma _{bb}^{{\rm in}}
\end{equation}
In the limit of large gain $G$ and for thermal fluctuations, it corresponds
to 
\begin{equation}
\hbar \left| \omega \right| \Sigma _{FF}^{{\rm out}}=k_{B}(\Theta
_{a}+\Theta _{b})
\end{equation}
When the temperatures are equal, this corresponds to a loss of $3dB$ in the
signal to noise ratio. This effect has been observed since the beginning of
radiowave communications. It also sets a limit in the number of repeaters in
optical fiber communications\cite{Gordon62,Takahasi65}.

Most practical applications of amplifiers in measurements involve ideal
operational amplifiers operating in the limits of infinite gain, infinite
input impedance and null output impedance. In order to deal with the
pathologies that could arise in such a system, we consider that it operates
with a feedback loop which fixes its effective gain and effective impedances 
\cite{Courty99}.

\begin{figure}[htb]
\centerline{\psfig{figure=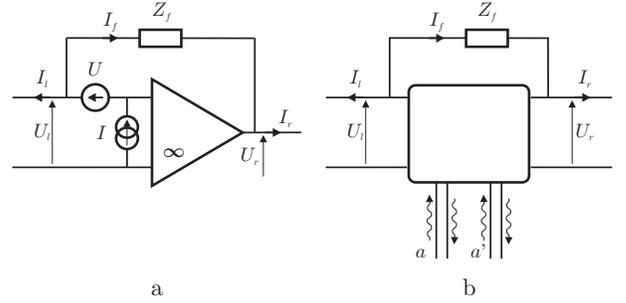,width=8cm}}
\vspace{2mm}
\caption{Representation of the ideal operational amplifier working in the
limit of infinite gain with a reactive feedback $Z_{f}$. (a) The noise
sources are described as a current generator $I$ and a voltage generator $U$%
. (b) The amplifier is represented with a left (input) port $l$ and a right
(output) port $r$ and the noise sources are modeled as input fields in the
two noise lines $a$ and $a^{\prime}$}
\label{Figure3}
\end{figure}

We first analyze the amplifier as depicted on figure \ref{Figure3}a where
the noise sources are represented as a current generator $I$ and a voltage
generator $U$. By coupling two coaxial lines denoted $l$ and $r$
respectively on the left port and the right port of the amplifier, one
realizes a measurement model. The left line comes from a monitored
electrical system so that the inward field $l^{{\rm in}}$ plays the role of
the signal to be measured. Meanwhile, the right line goes to an electrical
meter{\it \ }so that the outward field $r^{{\rm out}}$ plays the role of the
meter readout. In connection with the discussions of Quantum Non Demolition
measurements \cite{Braginsky92,Grangier92}, $l^{{\rm out}}$ appears as the
back-action field sent back to the monitored system and $r^{{\rm in}}$
represents the fluctuations coming from the readout line. A reactive
impedance $Z_{f}$ acts as feedback for the amplifier.

We now present the electrical equations associated with this measurement
device. We first write the characteristic relations between the voltages and
currents 
\begin{eqnarray}
U &=&U_{l}=U_{r}+Z_{f}I_{f}  \nonumber \\
I &=&I_{l}+I_{f}  \label{amplifier}
\end{eqnarray}
Here, $U_{p}$ and $I_{p}$ are the voltage and current at the port $p$, i.e.
at the end of the line $p=l$ or $r$, while $U$ and $I$ are the voltage and
current noise generators associated with the operational amplifier itself
(see Fig.1). $Z_{f}$ is the impedance feedback. All equations are implicitly
written in the frequency representation and the impedances are functions of
frequency. Equations (\ref{amplifier}) take a simple form because of the
limits of infinite gain, infinite input impedance and null output impedance
assumed for the ideal operational amplifier. We also suppose that the fields
incoming through the various ports are uncorrelated with each other as well
as with amplifier noises.

As already emphasized, the output fields $p^{{\rm out}}$ obey the
commutation relations (\ref{commutfree}) of free fields. To make this
property explicit, we use the characteristic equations (\ref{amplifier},\ref
{defnyquist} ) associated with the amplifier and the lines to rewrite the
output fields $l^{{\rm out}}$ and $r^{{\rm out}}$ in terms of input fields $%
l^{{\rm in}}$, $r^{{\rm in}}$ and of amplifier noise sources $U$ and $I$ 
\begin{eqnarray}
l^{{\rm out}} &=&-l^{{\rm in}}+\sqrt{\frac{2}{\hbar \left| \omega \right|
R_{l}}}U  \nonumber \\
r^{{\rm out}} &=&-r^{{\rm in}}-2\frac{Z_{f}}{\sqrt{R_{r}R_{l}}}l^{{\rm in}} 
\nonumber \\
&&+\sqrt{\frac{2}{\hbar \left| \omega \right| R_{r}}}\left( \frac{R_{l}+Z_{f}%
}{R_{l}}U-Z_{f}I\right)   \label{inout}
\end{eqnarray}
We then deduce from (\ref{inout}) that the voltage and current fluctuations $%
U$ and $I$ obey the following commutation relations 
\begin{eqnarray}
\left[ U\left[ \omega \right] ,U\left[ \omega ^{\prime }\right] \right]  &=&%
\left[ I\left[ \omega \right] ,I\left[ \omega ^{\prime }\right] \right] =0 
\nonumber \\
\left[ U\left[ \omega \right] ,I\left[ \omega ^{\prime }\right] \right] 
&=&2\pi \ \hbar \omega \ \delta \left( \omega +\omega ^{\prime }\right) 
\label{UU}
\end{eqnarray}
Hence, voltage and current fluctuations verify Heisenberg inequalities which
determine the ultimate performance of the ideal operational amplifier used
as a measurement device \cite{Courty99}.

To push this analysis further it is worth introducing new quantities $a^{%
{\rm in}}$ and $a^{\prime {\rm in}}$ as linear combinations of the noises $U$
and $I$ depending on a factor $R$ having the dimension of an impedance

\begin{eqnarray}
U\left[ \omega \right] &=&\sqrt{2\hbar \left| \omega \right| R}\left( a^{%
{\rm in}}\left[ \omega \right] -a^{\prime {\rm in}}\left[ -\omega \right]
\right)  \nonumber \\
I\left[ \omega \right] &=&\sqrt{\frac{2\hbar \left| \omega \right| }{R}}%
\left( a^{{\rm in}}\left[ \omega \right] +a^{\prime {\rm in}}\left[ -\omega %
\right] \right)  \label{deflignes}
\end{eqnarray}
For an arbitrary value of $R$, the quantities $a^{{\rm in}}$ and $a^{\prime 
{\rm in}}$ satisfy the free field commutation relations. In other words, the
voltage and current noises associated with the amplifier may be replaced by
the coupling to $2$ further lines $a$ and $a^{\prime }$ and the presence of
amplification requires a conjugation of fluctuations coming in one of these
two lines. This representation of the amplifier as a quantum network is
depicted on figure \ref{Figure3}b.

We may then fix the parameter $R$ to a value $R_{a}$ chosen so that the
fluctuations $a^{{\rm in}}$ and $a^{\prime {\rm in}}$ are uncorrelated. This
specific value is determined by the ratio between voltage and current noise
spectra 
\begin{equation}
R_{a}=\sqrt{\frac{\sigma _{UU}}{\sigma _{II}}}  \label{R0}
\end{equation}
The $2$ noise spectra $\sigma _{UU}$ and $\sigma _{II}$ are defined as
symmetric correlation functions. The fields $a^{{\rm in}}$ and $a^{\prime 
{\rm in}}$ are thus described by temperatures $T_{a}$ and $T_{a^{\prime }} $%
. We have assumed that these fluctuations are the same for all field
quadratures, i.e. that the amplifier noises are phase-insensitive. Although
this assumption is not mandatory for the forthcoming analysis, we also
consider for simplicity that the specific impedance $R_{a}$ is constant over
the spectral domain of interest.

\section{The cold damped accelerometer}

We come to the discussion of the ultimate performance of the cold damped
capacitive accelerometer designed for fundamental physics experiments in
space \cite{Grassia99}.

\begin{figure}[htb]
\centerline{\psfig{figure=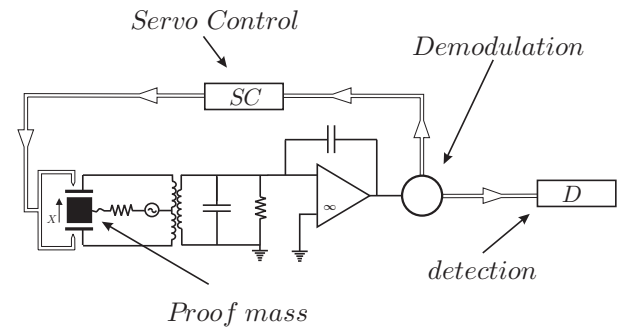,width=8cm}}
\vspace{2mm}
\caption{Scheme of the capacitive sensor. The proof mass is placed between
two electrodes formed by the inner walls of the accelerometer cage. The
position dependent capacitances are polarized by an AC sinewave source which
induces a mean current at frequency $\protect\omega _{t}$ in the symmetrical
mode. The mass displacement is read as the current induced in the
antisymmetric mode. An additional capacitance is inserted to make the
antisymmetric mode resonant with $\protect\omega _{t}$. The signal is
detected after an ideal operational amplifier with capacitive feedback
followed by a synchronous demodulation. The impedance of the detection line
plays the role of a further resistance $R_{r}$. The detected signal then
feds the servo loop used to keep the mass centered with respect to the cage.}
\label{Figure4}
\end{figure}

The central element of the capacitive accelerometer is a parallelepipedic
proof mass placed inside a box. The walls of these box are electrodes
distant from the mass off a hundred micrometers. The proof mass is kept at
the center of the cage by an electrostatic suspension. Since a three
dimensional electrostatic suspension is instable, it is necessary to use an
active suspension.

In the cage reference frame, an acceleration is transformed in an inertial
force acting on the proof mass. The force necessary to compensate this
inertial force is measured. In fact, as in most ultrasensitive measurements,
the detected signal is the error signal used to compensate the effect of the
measured phenomenon.

The essential elements of the accelerometer are presented in figure \ref
{Figure4}. The proof mass and the cage form two condensators. Any mass
motion unbalances the differential detection bridge and provides the error
signal. In order to avoid low frequency electrical noise, the electrical
circuit is polarized with an AC\ voltage with a frequency of a hundred
kilohertz. After demodulation, this signal is used for detection and as an
error signal for a servo control loop which allows to keep the mass centered
in its cage.

Furthermore, the derivative of this signal provides a force proportional to
the mass velocity and simulates a friction force. This active friction is
called cold damping since it may be noiseless. More precisely, the effective
temperature of the fluctuations of this active friction is much lower than
the physical temperature of the device.

The detection is performed with the output detection signal $r_{1}^{{\rm out}%
}$. It is a linear combination of the external force $F_{ext}$ and of input
fields in the various noise lines. We normalize this expression so that the
coefficient of proportionality appearing in front of the external force $%
F_{ext}$ is reduced to unity. With this normalization, we obtain a force
estimator $\widehat{F}_{ext}$ which is just the sum of the true force $%
F_{ext}$ to be measured and of an equivalent input force noise. In the
absence of feedback, the force estimator reads \cite{Grassia99}: 
\begin{equation}
\widehat{F}_{ext}=F_{ext}+\sum_{\alpha }\mu _{\alpha }\alpha ^{{\rm in}}
\label{estimfree}
\end{equation}
where $\alpha ^{{\rm in}}$ denote the various input fields corresponding to
the active and passive elements in the accelerometer.

When the feedback is active, the servo loop efficiently maintains the mass
at its equilibrium position and the velocity is no longer affected by the
external force $F_{ext}$. The residual motion is interpreted as the
difference between the real velocity of the mass and the velocity measured
by the sensor. This means that the servo loop efficiently corrects the
motion of the mass except for the sensing error. However the sensitivity to
external force is still present in the correction signal. Quite remarkably,
in the limit of an infinite loop gain and with the same approximations as
above, the expression of the force estimator $\widehat{F}_{ext}$ is the same
as in the free case \cite{Grassia99}.

The added noise spectrum $\Sigma _{FF}$ is obtained as 
\[
\Sigma _{FF}=\sum_{\alpha }\left| \mu _{\alpha }\right| ^{2}\sigma _{\alpha
\alpha }^{{\rm in}} 
\]
We have evaluated the whole noise spectrum $\Sigma _{FF}$ for the specific
case of the instrument proposed for the $\mu $SCOPE space mission devoted to
the test of the equivalence principle. Some of the main parameters of this
system are listed below 
\begin{eqnarray}
M=0.27{\rm \ kg} &\qquad &H_{m}=1.3\times 10^{-5}{\rm \ kg\ s}^{-1} 
\nonumber \\
\frac{\Omega }{2\pi }\simeq 5\times 10^{-4}{\rm \ Hz} &\qquad &\frac{\omega
_{t}}{2\pi }\simeq 10^{5}{\rm \ Hz}  \nonumber \\
R_{a}=0.15\times 10^{6}\ \Omega &\qquad &\Theta _{a}=1.5{\rm \ K}
\label{parameters}
\end{eqnarray}
$M$ is the mass of the proof mass, $H_{m}$ is the residual mechanical
damping force, $\frac{\Omega }{2\pi }$ is the frequency of the measured
mechanical motion, $\frac{\omega _{t}}{2\pi }$ is the operating frequency of
the electrical detection circuit. $R_{a}$ and $\Theta _{a}$ are the
characteristic impedance and temperature of the amplifier.

In these conditions, the added noise spectrum is dominated by the mechanical
Langevin forces 
\begin{eqnarray}
\Sigma _{FF} &=&2H_{m}k_{B}\Theta _{m}  \nonumber \\
&=&1.1\times 10^{-25}\left( {\rm kg\ m\ s^{-2}}\right) ^{2}/{\rm Hz}
\end{eqnarray}
This corresponds to a sensitivity in acceleration 
\begin{equation}
\frac{\sqrt{\Sigma _{FF}}}{M}=1.2\times 10^{-12}{\rm \ m\ s^{-2}}/\sqrt{{\rm %
Hz}}
\end{equation}
Taking into account the integration time of the experiment, this leads to
the expected instrument performance corresponding to a test accuracy of $%
10^{-15}$.

In the present state-of-the-art instrument, the sensitivity is thus limited
by the residual mechanical Langevin forces. The latter are due to the
damping processes in the gold wire used to keep the proof mass at zero
voltage \cite{Willemenot97}. With such a configuration, the detection noise
is not a limiting factor. This is a remarkable result in a situation where
the effective damping induced through the servo loop is much more efficient
than the passive mechanical damping. This confirms the considerable interest
of the cold damping technique for high sensitivity measurement devices.

Future fundamental physics missions in space will require even better
sensitivities. To this aim, the wire will be removed and the charge of the
test mass will be controlled by other means, for example UV photoemission.
The mechanical Langevin noise will no longer be a limitation so that the
analysis of the ultimate detection noise will become crucial for the
optimization of the instrument performance. This also means that the
electromechanical design configuration will have to be reoptimized taking
into account the various noise sources associated with detection \cite
{Grassia99}.


\begin{references}
\bibitem{Bernard91}  A.\ Bernard and P. Touboul, {\it The GRADIO
accelerometer: design and development status}, Proc. ESA-NASA Workshop on
the Solid Earth Mission ARISTOTELES, Anacapri, Italy (1991).

\bibitem{Touboul92}  P. Touboul et al., {\it Continuation of the GRADIO
accelerometer predevelopment}, ONERA\ Final Report 51/6114PY, 62/6114PY
ESTEC Contract (1992, 1993).

\bibitem{Willemenot97}  E. Willemenot, {\it Pendule de torsion \`{a}
suspension \'{e}lectrostatique, tr\`{e}s hautes r\'{e}solutions des acc\'{e}l%
\'{e}rom\`{e}tres spatiaux pour la physique fondamentale}, Th\`{e}se de
Doctorat de l'universit\'{e} Paris 11 (1997).

\bibitem{Einstein05}  A. Einstein, {\it Annalen der Physik} {\bf 17} (1905)
549.

\bibitem{Nyquist28}  H. Nyquist, {\it Phys. Rev.} {\bf 32} (1928) 110.

\bibitem{Callen51}  H.B. Callen and T.A. Welton, {\it Phys. Rev.} {\bf 83}
(1951) 34.

\bibitem{Landau84}  L. Landau and E.M. Lifshitz, {\it ``Course of
Theoretical Physics: Statistical Physics Part 1'' }(Butterworth-Heinemann,
1980) ch. 12.

\bibitem{Kubo66}  R. Kubo, {\it Rep. Prog. Phys.} {\bf 29} (1966) 255{\it .}

\bibitem{Landau}  E.M. Lifshitz and L.P. Pitaevskii, {\it ``Landau and
Lifshitz, Course of Theoretical Physics, Statistical Physics Part 2''}
(Butterworth-Heinemann, 1980) ch. VIII.

\bibitem{Planck}  M. Planck M. 1900 {\it Verh. Deutsch. Phys. Ges. }{\bf 13 }%
138 (1911); W. Nernst {\it ibid.} {\bf 18 }83 (1916)

\bibitem{Meixner63}  J. Meixner, {\it J. Math. Phys.} {\bf 4} (1963) 154.

\bibitem{Yurke84}  B. Yurke and J.S. Denker, {\it Phys. Rev.} A {\bf 29}
(1984) 1419.

\bibitem{Gardiner88}  C.W.\ Gardiner, {\it IBM J. Res. Dev.} {\bf 32} (1988)
127.

\bibitem{Courty92}  J-M. Courty and S. Reynaud, {\it Phys. Rev.} {\bf A 46}
(1992) 2766.

\bibitem{Francesca98}  F. Grassia, {\it Fluctuations quantiques et
thermiques dans les transducteurs \'{e}lectrom\'{e}caniques}, Th\`{e}se de
Doctorat de l'Universit\'{e} Pierre et Marie Curie (1998).

\bibitem{Grassia99}  F.\ Grassia, J.M.\ Courty, S.\ Reynaud and P.\ Touboul,
Eur.\ Phys.\ J.\ D {\bf 8}, 101, quant-ph/9904073

\bibitem{Feldmann}  M. Feldmann, {\it Th\'{e}orie des r\'{e}seaux et syst%
\`{e}mes lin\'{e}aires,} (Eyrolles 1986)

\bibitem{Heffner62}  H. Heffner, {\it Proc IRE} {\bf 50} (1962) 1604.

\bibitem{Haus62}  H.A.\ Haus and\ J.A.\ Mullen, {\it Phys. Rev.} {\bf 128}
(1962) 2407.

\bibitem{Gordon63}  J.P.\ Gordon, L.R. Walker and W.H. Louisell, {\it Phys.
Rev.} {\bf 130} (1963) 806.

\bibitem{Caves82}  C.M. Caves, {\it Phys. Rev.} {\bf D26} (1982) 1817.

\bibitem{Loudon84}  R.\ Loudon and T.J.\ Shephered, {\it Optica Acta} {\bf %
31 } (1984) 1243.

\bibitem{Gordon62}  J.P.\ Gordon, {\it Proc. IRE} (1962) 1898.

\bibitem{Takahasi65}  H.\ Takahasi, in {\it ``Advances in Communication
Systems''} ed. A.V. Balakrishnan (Academic, 1965) 227.

\bibitem{Courty99}  J-M. Courty, F. Grassia and S. Reynaud, {\it Europhys.
Lett}, {\bf 46} (1), pp. 31-37 (1999) quant-ph/9811062.

\bibitem{Braginsky92}  V.B. Braginsky and F.Ya. Khalili, ``{\it Quantum
Measurement}'' (Cambridge University Press, 1992).

\bibitem{Grangier92}  P. Grangier, J.M. Courty and S. Reynaud, {\it Opt.
Comm.} {\bf 89} (1992) 99.
\end{references}
\end{document}